\documentclass{ws-procs9x6}

\begin{document}

\title{Nuclear matter, nuclei, and neutron stars in hadron and quark-hadron models}

\author{S. Schramm}

\address{Frankfurt Institute for Advanced Studies, Goethe University,\\
Frankfurt, D-60438, Germany\\
$^*$E-mail: schramm@fias.uni-frankfurt.de
}
\author{V. Dexheimer}

\address{Gettysburg College,\\
300 North Washington Street
Gettysburg, PA 17325
USA\\
E-mail: vantoche@gettysburg.edu}

\author{R. Negreiros}

\address{Frankfurt Institute for Advanced Studies, Goethe University,\\
Frankfurt, D-60438, Germany\\
$^*$E-mail: negreiros@fias.uni-frankfurt.de
}

\author{T. Sch\"urhoff}

\address{Frankfurt Institute for Advanced Studies, Goethe University,\\
Frankfurt, D-60438, Germany\\
$^*$E-mail: schuerhoff@th.physik.uni-frankfurt.de
}

\begin{abstract}
We develop a unified model of hadrons and quarks. Within this approach we investigate the phase structure of the model 
as function of temperature and chemical potential. Computing the equation of state of cold matter we determine neutron and
hybrid star masses and radii. In an extension of the investigation we consider the cooling behavior of the compact stars
and derive a general relation between the star's mass and rotation and its cooling behavior.
Finally we study the effect of $\Delta$ resonances for star matter, especially with respect to possible solutions of stars
with small radii.   
\end{abstract}


\bodymatter

\section{Introduction}

A central topic in modern nuclear physics is the study of the general behavior of strongly interacting matter under extreme conditions.
One expects a transition from the effective degrees of freedom in vacuum and ground state matter, hadrons, to a phase of quarks and gluons at high temperatures
and/or densities. This can be described by the restoration of chiral symmetry and the deconfinement transition. The actual structure of the transition is still unknown, perhaps with the exception in the case of vanishing chemical potential, where lattice results show a cross-over transition. Some lattice results also point to a critical end-point of a first-order transition line in temperature and chemical potential \cite{fodor}.
In order to describe the matter, a main challenge for theory is to develop models that can cover all the ranges of temperatures and densities. This requires a model that is able to reproduce both nuclear matter ground state properties and the quark and gluon phase beyond the phase transition.
In the following we develop an effective theoretical description along these lines. Results for neutron star properties and their
cooling behavior within this approach will be presented. In addition we study the influence of the $\Delta$ baryons on stellar properties.

\section{Effective Chiral Model}

In order to describe hadronic physics as well as quark degrees of freedom we couple an effective chiral flavor-SU(3) model for hadrons \cite{chiral1} with a modified version of the so-called
PNJL quark model. The model includes the lowest SU(3) multiplets of hadrons and quarks. Baryonic and constitutent quark vacuum masses are generated via spontaneous
symmetry breaking by the coupling of the particles to the scalar fields, which attain non-zero vacuum expectation values.
Explicitly the Lagrangian reads  \cite{eu}:
\begin{eqnarray}
&L = L_{Kin}+L_{Int}+L_{Self}+L_{SB}-U,&
\end{eqnarray}
Here, the interaction term between baryons, quarks and mesons is given by
\begin{equation}
L_{Int}=-\sum_i \bar{\psi_i}[\gamma_0(g_{i\omega}\omega+g_{i\phi}\phi+g_{i\rho}\tau_3\rho)+M_i^*]\psi_i,\nonumber
\end{equation}
with the effective baryon and quark mass $M_i^*$
\begin{equation}
M_{i}^*=g_{i\sigma}\sigma+g_{i\delta}\tau_3\delta+g_{i\zeta}\zeta+\delta M_{i}~.
\end{equation}
$\delta M_{i}$ is a small explicit mass term. The self interactions of the scalar fields, generating the spontaneous symmetry breaking,
and of the vector fields are contained in
\begin{eqnarray}
&L_{Self}=-\frac{1}{2}(m_\omega^2\omega^2+m_\rho^2\rho^2+m_\phi^2\phi^2)\nonumber&\\&
+g_4\left(\omega^4+\frac{\phi^4}{4}+3\omega^2\phi^2+\frac{4\omega^3\phi}{\sqrt{2}}+\frac{2\omega\phi^3}{\sqrt{2}}\right)\nonumber&\\&+k_0(\sigma^2+\zeta^2+\delta^2)+k_1(\sigma^2+\zeta^2+\delta^2)^2&\nonumber\\&+k_2\left(\frac{\sigma^4}{2}+\frac{\delta^4}{2}
+3\sigma^2\delta^2+\zeta^4\right)
+k_3(\sigma^2-\delta^2)\zeta&\nonumber\\&+k_4\ \ \ln{\frac{(\sigma^2-\delta^2)\zeta}{\sigma_0^2\zeta_0}}.&
\end{eqnarray}
Finally, the explicit symmetry breaking term $L_{SB}$ has the form
\begin{equation}
L_{SB}= m_\pi^2 f_\pi\sigma+\left(\sqrt{2}m_k^ 2f_k-\frac{1}{\sqrt{2}}m_\pi^ 2 f_\pi\right)\zeta~.
\end{equation}
The mesons included are
the vector isoscalar fields $\omega$ and $\phi$, the vector isovector $\rho$,
the scalar isoscalars $\sigma$ and $\zeta$ (non-strange and strange quark-antiquark state, respectively) and  the scalar-isovector state  $\delta$.
The isovector mesons affect isospin asymmetric matter, which makes them relevant for the study of compact stars.
The various couplings are fitted to reproduce the vacuum masses of the baryons and mesons, nuclear saturation properties (density $\rho_0=0.15$ fm$^{-3}$, binding energy per nucleon $B/A=-16.00$ MeV, nucleon effective mass $M^*_N=0,67$ $M_N$, compressibility $K=297.32$ MeV), asymmetry energy ($E_{sym}=32.50$ MeV), and reasonable values for the hyperon potentials ($U_\Lambda=-28.00$ MeV, $U_\Sigma=5.35$ MeV, $U_\Xi=-18.36$ MeV).  The numerical values of the couplings can be found in Ref. \refcite{chiral1,eu2,eu}.

In order to be able to study the transition from hadronic to quark degrees of freedom the model incorporates a Polyakov loop field $\Phi$ in analogy
to the PNL models for quarks \cite{PNJL,Ratti1}. The field has an effective potential of the form
\begin{eqnarray}
&U=(a_0T^4+a_1\mu^4+a_2T^2\mu^2)\Phi^2&\nonumber\\&+a_3T_0^4\log{(1-6\Phi^2+8\Phi^3-3\Phi^4)}.&
\end{eqnarray}
depending on the temperature $T$ and the baryochemical potential $\mu$. The $\mu$ independent terms have been fitted, so that thermodynamical
quantities of quenched lattice calculations agree well with the model results over a wide range of temperature. The other two parameters of the potential
have been adjusted in order to obtain a phase diagram that has a first-order phase transitions at high densities which ends in a critical end-point at
a temperature $T$ and chemical potential $\mu$ in agreement with results from lattice calculations \cite{fodor}.
In order to suppress quarks at low densities and temperatures and hadrons in the deconfined phase an 
efffective mass shift is introduced of the form \cite{eu2}:
\begin{equation}
\delta M_{B}^* = g_{B\Phi} \Phi^2~~,~~
\delta M_{q}^*= g_{q\Phi}(1-\Phi)~.
\end{equation}

\section{Results}

In an earlier investigation, where we fine-tuned the couplings to properties of finite nuclei, we computed nuclei in a two-dimensional calculation including nuclear deformation \cite{def}~. The corresponding equations of motion for the mesonic fields as well as the Dirac equation for the nucleons were
solved on a spatial grid. Computing all nuclei with known binding energies an overall accuracy of the energies could be achieved with an average 
error for nuclei with 50 or more nucleons of
\begin{equation}
\epsilon_E (A > 50)  \approx  0.21 \%~~.
\end{equation} 
This result compares very well with standard relativistic and non-relativistic mean-field nuclear structure models.
As can be seen in Fig. 1 magic numbers can be reproduced as well as the measured deformations of many nuclei. Thus  this shows that a more general hadronic model approach
can be used to perform competitive nuclear structure calculations.

\begin{figure}
\begin{center}
\psfig{file=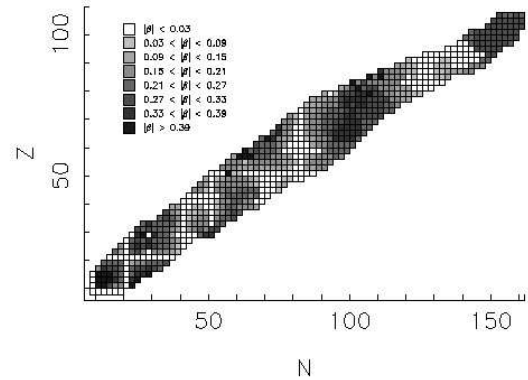,height=5cm}
\end{center}
\caption{\label{deformed}Nuclear chart as calculated in the effective chiral model. The gray level encodes the deformation (absolute value) of the corresponding
nucleus. One can clearly observe lines of spherical nuclei corresponding to magic numbers.}
\end{figure}

Investigating the full model and tuning the couplings in order to reproduce high-mass neutron stars as well as having reasonable nuclear ground state values 
\cite{eu} we minimize the grand canonical potential in mean-field approximation following from the Lagrangian discussed in the previous section \cite{eu}.
Figure 2 shows the resulting phase transition curve for symmetric nuclear matter and star matter in $\beta$ equilibrium. The critical end-point occurs
at $T_c \approx 167$~MeV and $\mu_c \approx 354$~MeV in agreement with Ref. \refcite{fodor}. With our choice of parameters the phase transition to quarks takes place at 4 times ground state density for vanishing temperature.
Since the model features a realistic nuclear ground state the figure also shows a quantitatively reasonable first-order liquid gas phase transition.

\begin{figure}
\begin{center}
\psfig{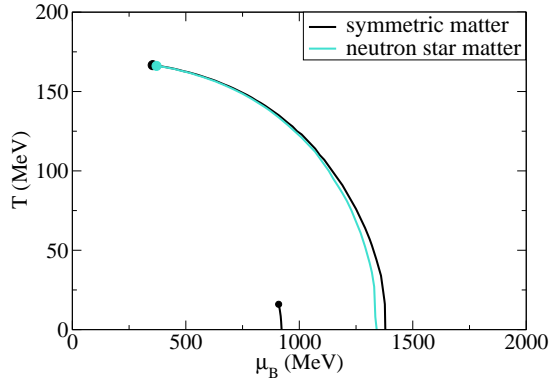}
\end{center}
\caption{\label{phase}Phase diagram as function of temperature and baryon chemical potential. The first order phase transition stops at a critical end point as suggested in Ref.  \refcite{fodor}. At $T = 0$ the transition takes place at 4 times nuclear matter saturation density. The phase transition for symmetric and star matter is shown.
}
\end{figure}

Computing the resulting field values one can determine energy density and pressure for stellar matter. Using the equation of state we solve the
Tolman-Oppenheimer-Volkov equations for non-rotating, spherical neutron stars \cite{tov1,tov2}. The resulting masses and radii of the stars are
shown in Fig. 3.  Excluding quarks the maximum attainable mass is about 2.1 solar masses. Taking quarks into account the maximum mass 
drops to $\approx 2$ solar masses. Note that, when quarks are populated, the region of stable stars is cut off as the equation of state in this regime
becomes softer and cannot support stable high-mass star solutions any more. This is true for a Maxwell construction of the phase transtion assuming local charge neutrality of the matter. In a Gibbs construction, which only assumes global charge neutrality, allowing for a mixture of quark and baryonic phases with 
opposite charge, this sharp cut off is softened as can be seen in the inset of the figure. As a result at the highest star masses a core of about 2 km size develops that contains a mixture of quarks and baryons. This can also be observed in Fig. 4, which shows the baryon number densities of baryons and quarks as function of
chemical potential. Around the phase transition, the plot shows this mixture of states. 
\begin{figure}
\begin{center}
\psfig{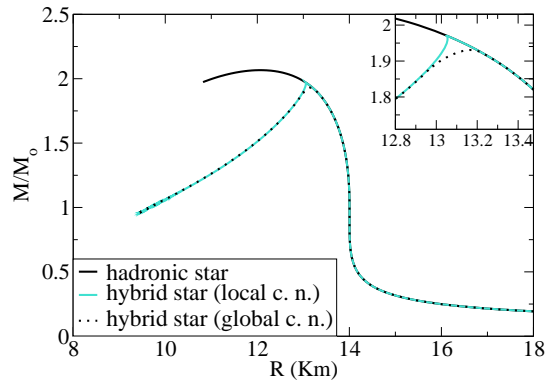}
\end{center}
\caption{\label{stars}Masses and radii of neutron stars. The results for purely baryonic stars and for stars including the quark phase are shown.
The inset illustrates the difference between a Maxwell and a Gibbs construction of the hadron-quark phase transition.
}
\end{figure}

\begin{figure}
\begin{center}
\psfig{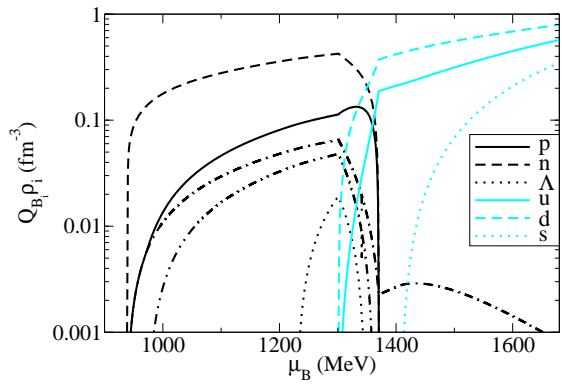}
\end{center}
\caption{\label{pop}Baryon number densities of particles for star matter as function of baryon chemical potential. The mixture of baryons and quarks results from the Gibbs construction of the first order phase transition assuming a mixed phase with global charge conservation.}
\end{figure}

One of the most interesting observables concerning neutron stars is their temperature compared to an age estimate of the star, defining the cooling behavior of the neutron star. To study this feature we make use of the equation of state of the hadron-quark model for stellar matter as described before. 
Incorporating this equation of state in a cooling simulation of the star we compute the cooling behavior of stars with different masses \cite{cooling}. Figure 5 shows the result of the calculation. Compared to the measurements also shown in the figure, one can conclude that a good agreement with data can be achieved if the corresponding (unknown) star masses are roughly
below 1.2 solar masses. For higher masses the model predicts much faster cooling rates. The main reason for this behavior is the on-set
of direct Urca processes for heavy stars. The most important processes are
\begin{equation}
n \rightarrow p + e^- + \bar{\nu}_e ~~~~,~~~ p + e^- \rightarrow  n + \nu_e ~~.
\end{equation}
These processes generate fast cooling via neutrino emission.
In a more general analysis we take into account rotation of the star. As rotation reduces the central energy density of the star also the threshold
of the direct Urca process shifts to higher masses. The result of this analysis is shown in Fig. 6. Here one can read off the ranges of frequencies and masses
of the stars that allow for the direct Urca process to take place, which leads to fast cooling of the star.
\begin{figure}
\begin{center}
\psfig{file=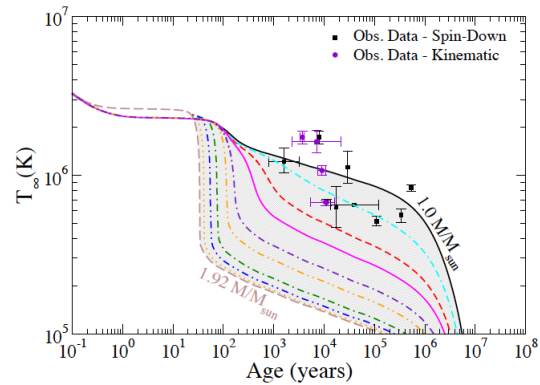,height=5cm}
\end{center}
\caption{Cooling behavior as function of time for compact stars as calculated in this model. Observational data are also shown. One can observe a good agreement with the data assuming low star masses below 1.2 solar masses (assuming non-rotating stars). 
For higher masses the direct URCA process leads to enhanced cooling.}
\end{figure}
\begin{figure}
\begin{center}
\psfig{file=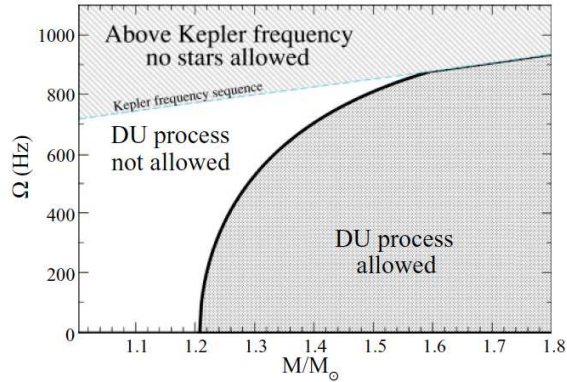,height=5cm}
\end{center}
\caption{The figure illustrates the range of star masses and rotation frequencies for which the Urca process takes place (DU process allowed).
The shaded region at high frequencies is not physically realized as the stars would be instable with respect to mass shedding.} 
\end{figure}

In a separate investigation concentrating on a purely baryonic star, for simplicity, we investigate the possibility of the existence of neutron stars with rather small radii around 10 km and below, as has been suggested in an analysis of masses and radii of three neutron stars discussed in Ref. \refcite{oezel}.
Here we also include the spin-3/2 baryonic multiplet as degrees of freedom as has already been studied in Ref. \refcite{eu2}.
As with other baryons the couplings of the particle to the scalar fields are fixed by reproducing baryonic vaccum masses.
The vector interaction strength is less clear. Using vector dominance arguments a natural choice is to take the same value as in the case of
the corresponding baryonic octet state with the same number of light and strange quarks. Starting from this assumption we allow for a small reduction from this canonical value \cite{delta}. The resulting star masses and radii are shown in Figure 7. One can see that for a moderate change in the vector strength of about 10 percent 
($r_V = 0.9$) one obtains good agreement with the values obtained in the analysis of the observational data. Thus it is possible to reproduce stars with relatively small radii in a purely baryonic model without introducing a more exotic solution like a quark star.
The population densities of the various baryons as function of chemical potential are shown in Fig. 8. It is interesting to observe a very different
structure compared to calculations not including the $\Delta$ resonances. In fact, beyond 4 times nuclear ground state densities the stellar matter is
dominated by  $\Delta$'s.

\begin{figure}
\begin{center}
\psfig{file=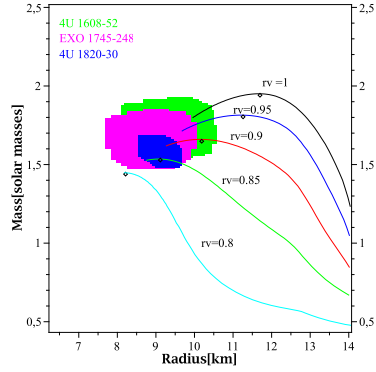,height=5cm}
\end{center}
\caption{Neutron star masses and corresponding radii. The strength of the vector-interaction of the Delta resonance is varied. Also shown are the results of an analysis of three observed neutron stars \cite{oezel}. $r_v$ denotes the value of the vector coupling of the $\Delta$ relative to the nucleonic value, 
i.e. $r_v = g_{\Delta\omega}/g_{N\omega}$. The model calculation can reproduce the observational data for a relative vector strength
$r_v = 0.9$. }
\end{figure}
\begin{figure}
\begin{center}
\psfig{file=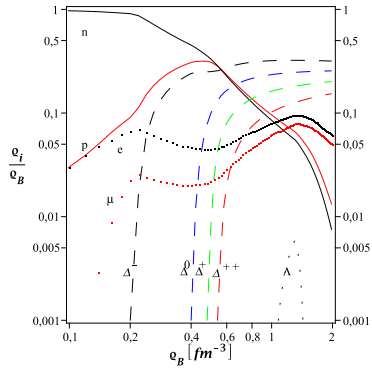,height=5cm}
\end{center}
\caption{Particle number densities as function of density including $\Delta$ resonances. One can observe that the $\Delta$s dominate other particles at densities $\rho > 4 \rho_o$.}
\end{figure}

\section{Summary}

We developed a unified hadron-quark model with the correct degrees of freedom on both sides of the QCD phase transition.
Within this approach we were able to reproduce a phase transition line with a critical end point for values of temperature
and chemical potential that were predicted by some lattice gauge calculations. Using these model parameters we studied neutron star
properties and found heavy stars with maximum mass values of around 2 solar masses. The core of the star contained a mixture
of baryons and quarks. Including a study of the cooling of the star we observed agreement with data for low-mass stars. 
In the case of high masses the direct Urca process led to fast cooling rates. We determine the general  parameter region of stars
with respect to rotational frequency and mass that did not allow for the Urca process to take place suppressing fast cooling of the stars.

In an investigation of purely baryonic stellar matter but taking into account the baryonic spin-3/2 multiplet, in particular the $\Delta$ resonances, we showed
that one is able to reproduce stars with small radii, which were suggested by the analysis of observational data. This was achieved by a slight tuning of the vector coupling strength of the spin-3/2 particles without the need for exotic particles.

\bibliographystyle{ws-procs9x6}

\end{document}